\documentclass[10pt,twocolumn,twoside]{article}

\usepackage{geometry}
\usepackage{graphics}
\usepackage{setspace}
\usepackage{graphicx}
\usepackage{amsmath}
\usepackage{amssymb}
\usepackage[T1]{fontenc}
\usepackage[scaled=0.9]{helvet}
\usepackage{sfmath}
\usepackage{hyperref}

\def\upd{\textup{d}}

\begin{document}

\title{\huge\bf Simple rules govern finite-size effects in scale-free networks\thanks{Published in \href{http://iopscience.iop.org/0295-5075/95/3/38002}{Europhysics Letters},
doi: \href{http://dx.doi.org/10.1209/0295-5075/95/38002}{10.1209/0295-5075/95/38002}}}

\author{
{Sara Cuenda$^{1,2,}$}\thanks{email: sara.cuenda@uam.es }
\,and\,
{Juan A. Crespo$^{1,}$}\thanks{email: juan.crespo@uam.es}\\
  \small $^{1}$ Departamento de Econom\'{\i}a Cuantitativa\\[-0.8ex]
  \small Facultad de CC. Econ\'omicas y Empresariales, Universidad Aut\'onoma de Madrid.\\
  \small $^{2}$ Grupo Interdisciplinar de Sistemas Complejos (GISC).
}

\date{}

\maketitle
\thispagestyle{empty}

\abstract{
We give an intuitive though general explanation of the finite-size effect in scale-free networks in terms of the  degree distribution of the starting network. This result clarifies the relevance of the starting network in the final degree distribution. We use two different approaches: the deterministic mean-field approximation used by Barab\'asi and Albert (but taking into account the nodes of the starting network), and the probability distribution of the degree of each node, which considers the stochastic process. Numerical simulations show that the accuracy of the predictions of the mean-field approximation depend on the contribution of the dispersion in the final distribution. The results in terms of the probability distribution of the degree of each node are very accurate when compared to numerical simulations.  The analysis of the standard deviation of the degree distribution allows us to assess the influence of the starting core when fitting the model to real data.
}

\section*{\normalsize Introduction}
Power-laws are not a new issue in scientific literature. The emergence of the scale-free behavior in the degree distribution of the sizes of biological genera, incomes, words in a text, scientific citation, etc., has been widely studied (and several times re-invented) in the past century (see \cite{review} for an interesting review of this re-inventions). One of the most famous works in this subject nowadays is the one by Barab\'asi, Albert and Jeong (BA) \cite{barabasi}, in which they introduce the ``preferential attachment'' model in social networks such as the world wide web \cite{barabasi:science} and the network of movie actors \cite{barabasi:prl}. In this model, nodes would get new links from new nodes in the network with probability proportional to their degree. This way, nodes with high degree would be more likely to receive new links, a {\em rich gets richer} mechanism that would render a power-law distribution of the network degree. This model has been widely used by other researchers (see, for example, \cite{newman:review} for an exhaustive review of other models constructed thereafter, and references therein).

However, some investigations showed that the model of preferential attachment of BA would depart from the predicted power-law behavior in small networks \cite{dorogovtsev:size,kaprivsky:2002,waclaw:2007,bascompte:2005}. An influence of the initial nodes from which the network starts growing was acknowledged, but no general prediction of the  effect of these nodes on the degree distribution was made, except for particular cases \cite{dorogovtsev:size,kaprivsky:2002,waclaw:2007}. Although the attempts of studying this effect have been both numerical and theoretical, up to our knowledge there are no intuitive, general explanations of the process, nor predictions of the final degree distribution in these networks in the scientific literature yet.

In this letter we find a general, theoretical prediction of the final degree distribution of finite networks growing with preferential attachment in terms of the degree distribution of the starting network. We obtain an expression of the final distribution using two different approaches: the well-known, deterministic mean-field approximation (with the contribution of the nodes of the starting network), and the expected probability distribution of the degree of each node, which considers the stochastic process. The methods used are very simple and intuitive, and the numerical simulations support very well the theoretical results. One of our main findings is the relevance of the starting nodes of the network in the final degree distribution, which must be considered when fitting the model to real data. 

\section*{\normalsize Model definition}
The model on which we are going to focus is the original one introduced by BA \cite{barabasi}. In this model, at every time step, a new node arrives to the network and attaches to other nodes by $m$ {\em undirected} new links, the probability of any node in the network of gaining one of these new links being proportional to its degree. Notice that, in order for the process to be well defined, a starting network (or core) to which the first new node may link is needed. We will not allow multiple linking between two nodes, thus the size of the starting core must be of $m$ nodes, at least. We do not consider nodes with null degree since, in this model, these nodes would never get any links.

We define $t$ as the number of nodes at each time step and $t_0$ as the number of nodes in the starting core, with $t_0\ge m$. If a node has degree $k$ at time $t$, then the probability of this node gaining a new link when node $t+1$ arrives is, according to the model of BA \cite{barabasi}, $\pi_{k,t}=Ak$, where $A$ is a normalizing constant that must satisfy the expected number of new links in the network to be $m$, $\sum_{i=1}^t\pi_{k_i,t}=m$. This condition renders
\begin{equation}\label{eq:pi}
\pi_{k,t}=\frac{mk}{\sum_{i=1}^t k_i}.
\end{equation}

If the mean number of links in the starting core is $m_0$, then the total degree of the network at time $t$ is $\sum_{i=1}^t k_i=2[(m_0-m)t_0+mt]=2m(\mu t_0+t)$, with $\mu=(m_0/m-1)$. Notice that $\pi_{k,t}=k/(2(\mu t_0+t))$ is such that the dynamics of every node is independent from the rest of nodes in the network. Therefore, whatever the approach to simulate the dynamics of each node, we can use this result to calculate the degree distribution of the network for each $t\ge t_0$, knowing the initial distribution at $t_0$.

\section*{\normalsize Mean-field approximation}
We start with the mean-field approach followed by BA \cite{barabasi}, which consists in a continuum approximation in both degree and time in such a way that the rate of change of the degree of any node, $k$, equals its expected value, $\pi_{k,t}$:
\begin{equation}\label{eq:dkdt}
\frac{dk}{dt}=\frac{k}{2(\mu t_0+t)}.
\end{equation}
Integration of eq.~\eqref{eq:dkdt} renders the deterministic degree at time $t$ of a node that has degree $\kappa$ at time $\tau$, $k(t)=h(t;\kappa,\tau)$, with
\begin{equation}\label{eq:kt}
h(t;\kappa,\tau)=\kappa\sqrt{\frac{\mu t_0+t}{\mu t_0+\tau}}
\end{equation}
and $t,\tau\ge t_0$. From expression \eqref{eq:kt} it follows that, for fixed $t$ and $\tau$, $k(t)$ strictly increases with $\kappa$, and for fixed $t$ and $\kappa$, $k(t)$ strictly decreases with $\tau$. Usually, $\kappa=m$ is taken for all nodes, and the usual asymptotic, power-law behavior is obtained \cite{barabasi}. However, notice that this initial condition is only valid for the nodes added to the network. When we consider the case of the added nodes ($\tau>t_0$) separately from the case of the nodes of the starting core ($\tau=t_0$), the finite-size effect emerges.

Let $F_{m}(k,t)$ be the complementary, cumulative distribution of the degree of the network at time $t$ under the mean-field approximation. Thus $F_{m}(k,t)$ gives the portion of nodes at time $t$ with degree greater than or equal to $k$ for $t\ge t_0$. Let $F_0(k)$ be the corresponding degree distribution of the starting core at time $t_0$, thus $F_{m}(k,t_0)=F_0(k)$. We define $k_m(t)$ as the degree at time $t$ of the nodes that had degree $m$ at time $t_0$, i.e., $k_m(t)=h(t;m,t_0)$. Therefore, all the added nodes of the network must have degree smaller than $k_m(t)$ at time $t$. Similarly, we define $k_0(k,t)$ as the degree that should have a node at time $t_0$ in order to have degree $k$ at time $t$; from eq.~\eqref{eq:kt} it follows that $k_0(k,t)=h(t_0;k,t)$.

At time $t$, nodes with degree $k>k_m(t)$ cannot come from the added nodes, but from the nodes that in the starting core had degree $k_0(k,t)$, instead. Thus the portion of nodes with degree greater than or equal to $k$, with $k>k_m(t)$, is $(t_0/t)\,F_0(k_0(k,t))$. For $m< k\le k_m(t)$, the portion of nodes with degree greater than or equal to $k$ coming from the starting core are $(t_0/t)\,F_0(k_0(k,t))$, but also added nodes until time $\tau^\ast$, with $\tau^\ast$ such that $h(t;m,\tau^\ast)=k$, must be considered, rendering $(\tau^\ast-t_0)/t$. Finally, for $k\le m$ all added nodes have degree greater than or equal to $k$, and the contribution from the starting core is similar to the other ranges. Therefore, the final degree distribution at time $t$ is
\begin{equation}\label{eq:fkt}
F_{m}(k,t)=\left\{
\begin{array}{ll}
1-\frac{t_0}{t}\left[1-F_0(k_0(k,t))\right],
&\!\! k\le m, \\
\left(\mu\frac{t_0}{t}+1\right)\left(\frac{m}{k}\right)^{2}-\mu\frac{t_0}{t}
& \\
\,\,-\frac{t_0}{t}\left[1-F_0(k_0(k,t))\right],
&\!\!m<k\le k_m(t), \\
\frac{t_0}{t}F_0(k_0(k,t)),
&\!\! k>k_m(t).
\end{array}\right.
\end{equation}
The finite-size effect that makes expression \eqref{eq:fkt} to depart from the classical power-law comes from the starting core in terms of $(t_0/t)F_0(k_0(k,t))$ and from the finite number of added nodes, which yields the emergence of $k_m(t)$. Notice that these contributions vanish in the limit $t\rightarrow\infty$, where the usual asymptotic result is recovered.

Previous works of the finite-size effect \cite{dorogovtsev:size,kaprivsky:2002,waclaw:2007} studied the ratio between the actual final degree density distribution of the finite network and the asymptotic power-law of BA and found that, for networks growing from the same starting core, this ratio is
\begin{equation}\label{eq:prev}
\frac{f(k,t)}{f(k,t\rightarrow\infty)}=w(k/\sqrt{t}),
\end{equation}
where $w(x)$ is the cut-off function, and depends on the starting core used. From expression \eqref{eq:fkt} we can calculate such a ratio within the mean-field approximation, rendering
\begin{equation}\label{eq:w}
\frac{f_m(k,t)}{f_m(k,t\rightarrow\infty)}=\frac{\mu t_0+t}{t}u(k,t)
\end{equation}
where $f_m(k,t)=-\partial F_m(k,t)/\partial k$ is the density distribution in the mean-field approximation, $f_m(k,t\rightarrow\infty)=2m^2k^{-3}$ is the asymptotic density obtained with the BA methodology,
\begin{equation}
u(k,t)=\left\{
\begin{array}{ll}
1+g\left(k_0(k,t)\right),& m<k<k_m(t),\\
g\left(k_0(k,t)\right),& k>k_m(t),
\end{array}
\right.
\end{equation}
$g(k)=\frac{1}{2m^2(\mu +1)}k^3f_0(k)$ and $f_0(k)=-\upd F_0(k)/\upd k$ is the degree density distribution of the starting core. Noticing that $k_0(k,t)=k\sqrt{(\mu t_0 +t_0)/(\mu t_0+t)}$, expression \eqref{eq:w} resembles expression \eqref{eq:prev} when $\mu t_0/t\simeq0$.

Expressions \eqref{eq:fkt} and \eqref{eq:w} are obtained within the well-known mean-field approximation as a consequence of considering the starting core degree distribution in the calculation of $F_m(k,t)$. However, it is strictly deterministic: the degree distribution of the starting $t_0$ nodes at time $t$ is the same initial distribution at time $t_0$, stretched by a factor $\sqrt{(\mu t_0+t_0)/(\mu t_0+t)}$; no effect of the dispersion of the degree of nodes as a consequence of the stochastic process is considered at all. In the next section we will consider the full stochastic process when calculating the final degree distribution.

\section*{\normalsize Probability distribution of the degree of a node}
Expression \eqref{eq:fkt} is the result of a counting procedure including the starting nodes within the deterministic mean-field approximation. However, a similar procedure with no further approximations can be followed in order to obtain the expected complementary, cumulative degree distribution of the network, $F_{e}(k,t)$. With mean-field, the stochastic dynamics of every node was approximated by the deterministic eq.~\eqref{eq:dkdt}; but the real stochastic process can be described by the probability distribution $P(k,t|\kappa,\tau)$, which represents the probability of a node having degree $k$ at time $t$ if it had degree $\kappa$ at time $\tau$, $t\ge \tau$. The probability of gaining a new link at each time step, $\pi_{k,t}$, relates to the probabilities $P(k,t|\kappa,\tau)$ and $P(k,t-1|\kappa,\tau)$ in the recurrence relation
\begin{equation}\label{eq:rec}
\begin{split}
P(k,t|\kappa,\tau)=&P(k,t-1|\kappa,\tau)(1-\pi_{k,t-1})\\
&+P(k-1,t-1|\kappa,\tau)\pi_{k-1,t-1}.
\end{split}
\end{equation}
\begin{figure}
\includegraphics[width=\columnwidth]{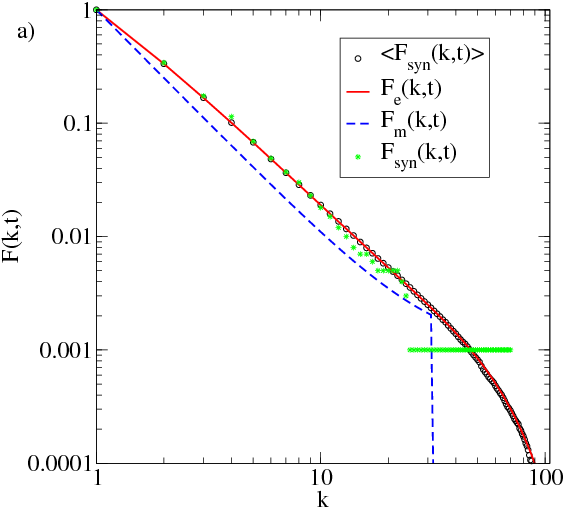}\\[0.8cm]
\includegraphics[width=\columnwidth]{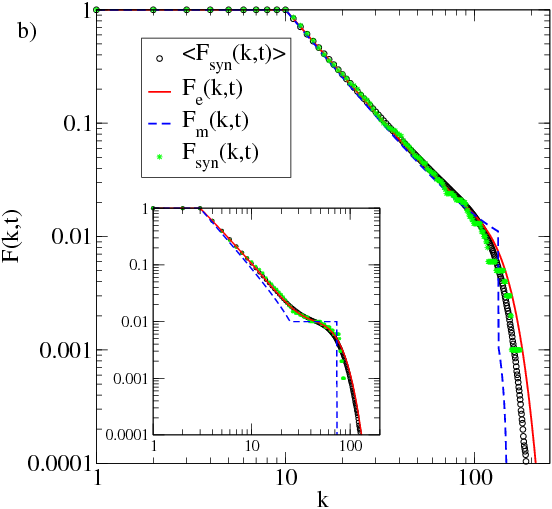}
\caption{\label{fig:1}
Complementary cumulative distributions versus degree in log-log scale for a network of $t=1000$ nodes. $\circ$: averaged distribution of 1000 synthetically generated networks, $\langle F_{syn}(k,t)\rangle$; $\ast$: final distribution of one of the synthetical networks, $F_{syn}(k,t)$; red, solid line: theoretical, expected distribution $F_e(k,t)$; blue, dashed line: theoretical mean-field approximation $F_m(k,t)$. The networks grow from an starting core with:
(a) only two linked nodes ($t_0=2$ nodes with degree distributed according to $\delta_{1,k}$), with $m=1$;
(b) $t_0=10$ nodes, with every node connected to the others (and therefore with a degree distributed according to $\delta_{9,k}$) and $m=10$ (inset: the same, for $m=3$).
}
\end{figure}

The number of links that a node may receive is constrained by the number of added nodes, $k-\kappa\le t-\tau$, and the degree of any node cannot decrease, so $k\ge \kappa$ for every $t\ge \tau$; thus $P(k,t|\kappa,\tau)=0$ for all the sets $(k,t)$ and $(\kappa,\tau)$ that do not fulfill these conditions. Therefore, $P(k,t|\kappa,\tau)$ can be numerically calculated from eq.~\eqref{eq:rec} using $P(\kappa,\tau|\kappa,\tau)=1$ as the initial condition.  Defining the probability distribution $f_{e}(k,t)$ as the expected probability of finding a node in the network with degree $k$ at time $t$, then the expression of $f_e(k,t)$ is
\begin{equation}\label{eq:fex}
f_{e}(k,t)=\sum_{\kappa=0}^k\frac{t_0}{t}f_0(\kappa)P(k,t|\kappa,t_0)+
\sum_{\tau=t_0+1}^t\frac{1}{t}P(k,t|m,\tau),
\end{equation}
where $f_0(\kappa)$ stands for the probability distribution of the degree of the starting core at time $t_0$, and thus $f_e(k,t_0)=f_0(k)$.
The first term in expression \eqref{eq:fex} comes from the nodes in the starting core: $\frac{t_0}{t}f_0(\kappa)$ stands for the expected fraction of nodes at time $t$ that had degree $\kappa$ in the starting core, and $P(k,t|\kappa,t_0)$ is the probability of these nodes of having degree $k$ at time $t$; the result is summed for all possible degrees that may affect $f_{e}(k,t)$. The second term refers to the added nodes: from the node added at time $t_0+1$ to the last node added at time $t$, the probability of these nodes having degree $k$ at time $t$ is summed and rescaled to the actual number of nodes, $t$. The cumulative distribution $F_{e}(k,t)$ can be calculated from \eqref{eq:fex} using the expression
\begin{equation}\label{eq:Fex}
F_{e}(k,t)=\sum_{j\ge k}f_{e}(j,t).
\end{equation}

\section*{\normalsize Numerical results}
In order to check the results of expressions \eqref{eq:fkt}, \eqref{eq:fex} and \eqref{eq:Fex} we need to simulate the stochastic process defined by expression \eqref{eq:pi}. However, for $m>1$, this process does not cover the space of probabilities of the whole system, i.e., the sum of the probabilities for all possible choices of the nodes that get a new link is not normalized to unity, and thus condition
\begin{equation}\label{eq:condpi}
\sum_{i_1=1}^t\sum_{i_2<i_1}\ldots\sum_{i_m<i_{m-1}}
\pi_{k_{i_1},t}\pi_{k_{i_2},t}\cdots\pi_{k_{i_m},t}=1
\end{equation}
is not fulfilled for $m>1$ (though it is for $m=1$). Therefore, for $m>1$, there is no such stochastic process.

Nonetheless, we can consider the following stochastic linking process when a new node arrives: the $m$ nodes are chosen sequentially, each one with probability $k/\sum_ik_i$, where the sum does not contain previously chosen nodes, and avoiding repetition. Expression \eqref{eq:pi} describes exactly this stochastic process for $m=1$, and it is a good approximation for $t\gg m$. For $t\gtrsim m$, there is an exclusion effect that makes the probability highly non-linear with respect to the degree (the case $t=m$, where all nodes should get a new link with probability equal to 1, shows the inaccuracy), and the model does not describe well the growing process in this regime. However, as the network grows, the model will eventually capture the stochastic dynamics of the nodes. As a result, for $t_0\gtrsim m$ and $t\gg m$, the dynamics of young nodes are well approximated by expression \eqref{eq:pi}, but the dynamics of old nodes may depart from that. These effects render a slight error in the prediction of the final distribution $F(k,t)$ for large values of $k$.

\begin{figure}
\includegraphics[width=\columnwidth]{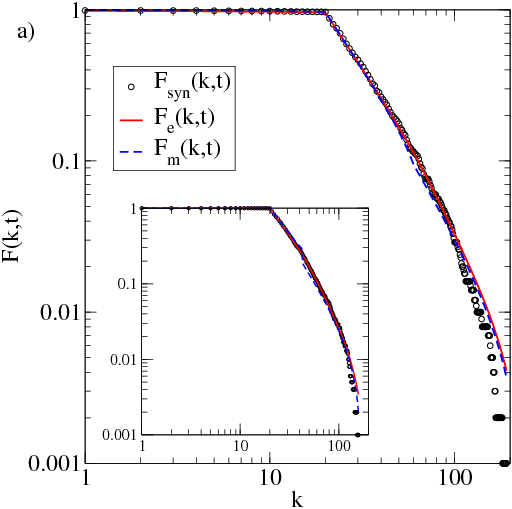}\\[8mm]
\includegraphics[width=\columnwidth]{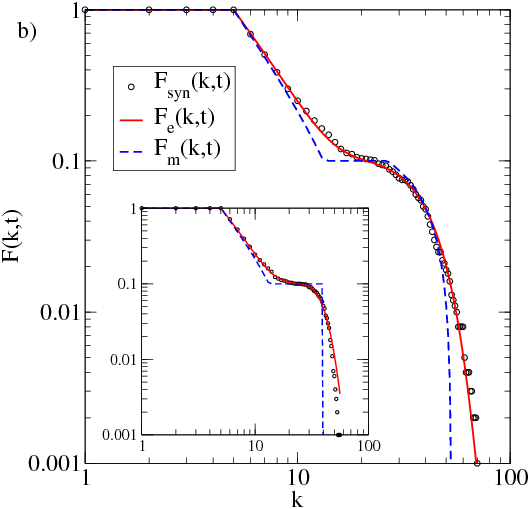}
\caption{\label{fig:2}
Complementary cumulative distributions versus degree in log-log scale for a network constructed from $t_0=100$ until $t=1000$ nodes. $\circ$: distribution of a synthetically generated network, $F_{syn}(k,t)$; red, solid line: theoretical, expected distribution $F_e(k,t)$; blue, dashed line: theoretical mean-field approximation $F_m(k,t)$, from an starting core with:
(a) degree distributed according to a $\delta_{20,k}$ probability distribution and $m=10$ new links per added node (inset: the same, but with a starting core with degree distributed according to $\delta_{5,k}$ and $m=20$);
(b) degree of the starting core distributed according to a uniform distribution between degree $10$ and $20$ and $m=5$ (inset: the same, for a uniform distribution between $0$ and $20$).
}
\end{figure}

Figures \ref{fig:1} and \ref{fig:2} show the agreement for different starting cores between the mean-field approximation $F_m(k,t)$, the expected $F_e(k,t)$ and synthetically generated complementary cumulative distributions, $F_{syn}(k,t)$. The growing process of fig.~\ref{fig:1}a has $m=1$, and therefore the result given by $F_e(k,t)$ is exact in this case. However, it is also the worst example for the mean-field approximation, $F_m(k,t)$, since the tail of the distribution comes from the dispersion of the degree of the starting nodes in the network; this is clear when comparing the tail of the distribution of the averaged simulations, $\langle F_{syn}(k,t)\rangle$, with one of them, $F_{syn}(k,t)$. In fig.~\ref{fig:1}b there is less dispersion, and the mean-field approximation works better. The expected distribution $F_e(k,t)$ departs slightly in the tail from $\langle F_{syn}(k,t)\rangle$ since the starting number of nodes is $t_0=10$ and the number of links per new node is $m=10$, but the result is not so bad since, as we add nodes to the network, the model improves its accuracy.

In fig.~\ref{fig:2} we see the effect in the final distribution of a poorly connected (fig.~\ref{fig:2}a) and a highly connected (fig.~\ref{fig:2}b) starting core compared to the number of links per new node, $m$. The heap that emerges in the tail of the latter comes from the degree of the starting nodes, higher than the initial degree of the added nodes. In the mean-field approximation, the position of the heap depends on $k_m(t)$ and $F_0(k_0(k,t))$. For poorly connected cores, $k_m(t)$ is very high compared to the degrees where the starting nodes contribute to $F_m(k,t)$ (notice the small perturbation of $F_m(k,t)$ around $k=50$ or $60$ in fig.~\ref{fig:2}a), and the tail of the network comes from the added nodes at every time step. On the contrary, for highly connected cores, the contribution of the starting nodes overtakes $k_m(t)$ and we see clearly the heap that they form. This analysis explains previous numerical results of the finite-size effect \cite{bascompte:2005}.

Figure \ref{fig:2}a is also an example the where mean-field gives a good approximation of the final distribution. This agreement comes from the low dispersion of the starting nodes in this case.

\begin{figure}
\includegraphics[width=\columnwidth]{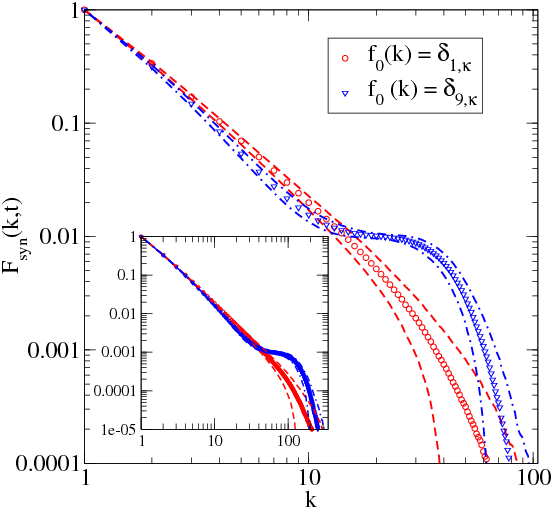}
\caption{\label{fig:sd}
Synthetically generated complementary cumulative distributions versus degree for a network of $t=1000$ nodes with a starting core of $t_0=10$ nodes with initial distributions $f_0(k)=\delta_{1,k}$ ($\circ$) and $f_0(k)=\delta_{9,k}$ ($\triangledown$). The distributions have been averaged over 1000 simulations, and the dashed lines correspond to the mean $\pm$ one standard deviation. The inset shows the final degree distributions when the network grows up to $t=10^4$ nodes.
}
\end{figure}

In figure \ref{fig:sd} the averaged final distributions and their standard deviations are represented for two different initial configurations and two different final network sizes. This plot shows the relevance of the differences that may arise in the final distribution depending on the starting core considered. The main differences arise for $F<t_0/t$, where the two averaged distributions behave completely differently and there is a region where the distributions do not overlap within the margin of each other's standard deviation. For $F>t_0/t$, the initial nodes make the slopes to be slightly different, too. As the inset shows, these effects remain even when the final network is ten times larger than the network of the main plot. Thus the finite-size effect is not embedded within the errors of the final distribution as the network grows. Such significantly different final distributions should give rise to different fitting results for the same empirical data. 

\section*{\normalsize Conclusions}
In this letter we have shown the basic mechanisms that lead to the finite-size effect in growing networks with preferential attachment, namely the nodes of the starting core from which the network starts growing and the dispersion in the degree of such nodes during the stochastic process. We have developed a general formalism that can be used with any deterministic or probabilistic approach of the dynamics of the degree of a node. In particular, we have shown the results with the mean-field approximation \eqref{eq:dkdt} and with the probability distribution $P_{k,t}(\kappa,\tau)$ defined in expression \eqref{eq:rec}. This methodology can be applied for any other approach to the dynamics of the degree of a node, though, as long as the degree distribution of the starting core of nodes from which the network starts its growing process, $F_0(k)$, is taken into account when calculating the distribution $F(k,t)$ of the final network.

We have shown that our results give an intuitive explanation of the apparently universal behavior of the cut-off function of the distribution observed in networks with different sizes when growing from the same starting core \cite{dorogovtsev:size,kaprivsky:2002,waclaw:2007} (see expressions \eqref{eq:prev} and \eqref{eq:w}). We have also explained previous results that studied the emergence of a heap in the final distribution \cite{bascompte:2005} in terms of the connectivity of the starting core compared to the number of links per new node, $m$, like the ones showed in fig.~\ref{fig:2}.

We have also shown that in the original model of BA \cite{barabasi} for $m>1$ the attachment process cannot be described with a probability of attachment to a node independent of that of other nodes. However, the probability given by expression \eqref{eq:pi} is a good approximation of the stochastic dynamics used in the simulations for $t\gg m$ when $m>1$. This must be considered when comparing the theoretical results obtained in this letter with numerical simulations, since the model of BA \cite{barabasi} (and, therefore, our theoretical results) are only accurate in this regime. For regimes with $t\gtrsim m$, the dynamics of the nodes would not be well described by expression \eqref{eq:pi} although, as we add nodes to the network, the approximation improves. This effect renders the differences observed in the tail of the distributions between the theoretical and the synthetical results in figs. \ref{fig:1} and \ref{fig:2} (specially in fig.~\ref{fig:1}b).

The methodology followed in this letter can also be applied to other growing network models where the dynamics of the degree of a single node can be well approximated by its own state, regardless of the state of other nodes or the degree distribution of the network. 
These models can lead in the asymptotic limit $t\rightarrow\infty$ to power-law-tailed distributions (see \cite{newman:review}) that are fitted against real data which is supposed to be modelled by this kind of growing mechanisms. However, the shape of the final distributions of these models may depend strongly on the initial configuration used, even for large networks, as shown in figure \ref{fig:sd}, where the final network is $10^2$ and $10^3$ times larger than the starting core. Clearly, the fitting results may differ significantly depending on the size and distributions of the intial cores of the models, and therefore the results presented in this letter should be considered.

Nowadays, dynamic networks are the hot topic in network investigation. Reaction kinetics on metabolic networks, spread of information or viruses in social networks\ldots~are examples of dominant issues in the latest scientific literature. But, quoting A.-L.~Bar\'abasi, ``to make progress in this direction, we need to tackle the next frontier, which is to understand the dynamics of the processes that take place on networks'' \cite{barabasi:decade}. We hope that this work may help in that understanding.

\section*{\normalsize Acknowledgements}
SC is partially supported by MEC grant FIS2006-01485. JAC is partially supported by MEC grant ECO2010-19596.


\begin{thebibliography}{1}
\expandafter\ifx\csname url\endcsname\relax\def\url#1{\texttt{#1}}\fi

\bibitem{review}
M.~Simkin and V.~Roychowdhury, ``Re-inventing willis'', \emph{Physics
  Reports} \textbf{502} p.1-35 (2011).

\bibitem{barabasi}
A.-L.~Barab\'asi, R.~Albert and H.~Jeong, ``Mean-field theory for scale-free random networks'', \emph{Phys.\ A} \textbf{272} p.173-187 (1999).

\bibitem{barabasi:science}
A.-L.~Barab\'asi, R.~Albert, H.~Jeong and G.~Bianconi, ``Powerlaw distribution of the world wide web'', \emph{Science} \textbf{287} p.2115 (2000).

\bibitem{barabasi:prl}
R. Albert and A.-L.~Barab\'asi, ``Dynamics of complex systems: scaling laws for the period of boolean networks'', \emph{Phys.\ Rev.\ Lett.} \textbf{84} p.5660 (2000).

\bibitem{newman:review}
M.~E.~J.~Newman, ``The structure and function of complex networks'', \emph{SIAM Rev.} \textbf{45} p.167 (2003).

\bibitem{dorogovtsev:size}
S.~N.~Dorogovtsev and J. F.~F.~Mendes, ``Size dependent degree distribution of a scale free growing networks'', \emph{Phys.\ Rev.\ E} \textbf{63} p.62101 (2001).

\bibitem{kaprivsky:2002}
P.~L.~Krapivsky and S.~Redner, ``Finiteness and fluctuations in growing networks'', \emph{J. Phys. A: Math. Gen.} \textbf{35} p.9517-34 (2002).

\bibitem{waclaw:2007}
B.~Waclaw and I.~M.~Sokolov, ``Finite-size effects in Barab\'{a}si-Albert growing networks'', \emph{Phys.\ Rev.\ E} \textbf{75} p.056114 (2007).

\bibitem{bascompte:2005}
P.~R.~Guimar\~{a}es et al., ``Random initial condition in small Barab\'{a}si-Albert networks and deviations from the scale-free behavior'', \emph{Phys.\ Rev.\ E} \textbf{71} p.37101 (2005).

\bibitem{barabasi:decade}
A.-L.~Barab\'asi, ``Scale-free networks: a decade and beyond'', \emph{Science} \textbf{325} p.412 (2009).

\end{thebibliography}

\end{document}